\documentclass{ptephy}
\usepackage{amsmath,times,bm}

\newcommand\hb[1]{\hat {\bm #1}}

\newcommand\cc{{\rm c.c.}}

\DeclareMathOperator\diag{diag}
\DeclareMathOperator\tr{tr}
\begin{document}
\title{Heat transport as torsional responses and Keldysh formalism in a curved spacetime}
\author{Atsuo Shitade}
\address{Department of Physics, Kyoto University, Kyoto 606-8502, Japan}
\begin{abstract}
  We revisit a theory of heat transport in the light of a gauge theory of gravity
  and find the proper heat current with a corresponding gauge field,
  which yields the natural definitions of the heat magnetization and the Kubo-formula contribution to the thermal conductivity as torsional responses.
  We also develop a general framework for calculating gravitational responses by combining the Keldysh and Cartan formalisms.
  By using this framework, we explicitly calculate these two quantities
  and reproduce the Wiedemann-Franz law for the thermal Hall conductivity in the clean and non-interacting case.
  Finally, we discuss an effective action for the quantized thermal Hall effect in $(2 + 1)$-D topological superconductors.
\end{abstract}
\subjectindex{I53}
\maketitle
\section{Introduction} \label{sec:intro}
Gravity sometimes appears in condensed-matter physics.
Fifty years ago, Luttinger introduced a gravitational potential to calculate heat transport coefficients~\cite{PhysRev.135.A1505}.
Recently, it was shown that the spin current can be generated by mechanical rotation~\cite{PhysRevLett.106.076601,PhysRevB.84.104410,PhysRevB.87.180402}.
Viscosity is a response to torsion, which is neglected in general relativity~\cite{PhysRevLett.107.075502,Hidaka01012013,PhysRevD.88.025040}.
Now gravity is not only a research interest in the fields of high-energy physics and cosmology
but is a powerful tool to describe many kinds of external fields in condensed-matter physics.
A unified description from the viewpoint of gravity brings about a deeper understanding of such gravitational responses.

Among these gravitational responses, heat transport is one of the most important phenomena.
Traditionally, charge and heat transport phenomena have been investigated within the semiclassical Boltzmann theory~\cite{9780306463389}.
One exception is the anomalous Hall effect in ferromagnetic metals~\cite{RevModPhys.82.1539}.
Since it is a complicated phenomenon involving the intrinsic mechanism due to the multiband Berry-phase effect and extrinsic mechanisms due to disorder,
a systematic perturbation theory with respect to disorder strength, e.g., the Keldysh formalism,
is necessary~\cite{PhysRevLett.97.126602,PhysRevB.77.165103,PhysRevB.78.041305,PhysRevB.79.195129}.
Experimentally, the anomalous thermal Hall effect (THE) is useful
for investigating effects of inelastic scattering on the anomalous Hall effect at finite temperature~\cite{PhysRevLett.100.016601,PhysRevB.79.100404,PhysRevB.81.054414}.
The THE was also used to detect the magnon Hall effect in ferromagnetic insulators~\cite{Onose16072010,PhysRevB.85.134411}.
To understand these phenomena, a systematic quantum-mechanical formula for the thermal Hall conductivity (THC) is highly desired.
One important point is that the Kubo formula alone is not sufficient and should be augmented with the heat magnetization (HM)
to avoid the unphysical divergence at zero temperature~\cite{0022-3719-10-12-021,PhysRevB.55.2344,PhysRevLett.107.236601}.
These previous theories are unsatisfactory
because the scaling relations on the charge and heat currents are assumed without any microscopic explanations
and, furthermore, it remains unclear how to apply them to disordered or interacting systems.

The HM also appears in the context of topological superconductors (TSCs)~\cite{RevModPhys.82.3045,RevModPhys.83.1057}.
Electromagnetic responses are not effective because the U($1$) gauge symmetry is broken.
Instead, the THC is known to be quantized in $(2 + 1)$-D TSCs~\cite{PhysRevB.61.10267,PhysRevLett.108.026802,JPSJ.82.023602}.
In $(3 + 1)$-D TSCs, it was theoretically proposed that
the HM is induced by a temperature gradient and the heat polarization by an angular velocity of rotation~\cite{PhysRevLett.108.026802}.
A possible question is what actions describe these topological phenomena.
Note that the spin-connection analogs of the Chern-Simons term and axion electrodynamics~\cite{PhysRevB.61.10267,PhysRevB.84.014527,PhysRevB.85.045104}
cannot do because these actions do not vanish at zero temperature.

Our main purposes are twofold.
One is to revisit the basics of heat transport from the gauge-theoretical viewpoint.
Compared to charge transport where the U($1$) gauge symmetry plays an important role,
heat transport has been less discussed regarding a symmetry and its gauge field.
As a result, the definition of the heat current has been controversial:
the product of the Hamiltonian and the velocity or that of the time derivative and the velocity.
Here we propose that the latter has a corresponding gauge field and is the proper definition.
We also find that a torsional magnetic field induced by this gauge field is conjugate to the HM.
In other words, the HM can be obtained by using a perturbation theory with respect to a torsional magnetic field.
We emphasize that this torsional magnetic field is
totally different from a gravito-magnetic field or an angular velocity discussed previously~\cite{PhysRevLett.107.236601,PhysRevB.85.045104,PhysRevLett.108.026802}.

The other is to establish a unified framework for calculating gravitational responses.
The gauge-covariant Keldysh formalism is
one of the most sophisticated perturbation theories with respect to electromagnetic fields~\cite{PTP.116.61,PTP.117.415,Sugimoto15062012}
and can be applied to insulators or metals with disorder or interactions at finite temperature.
By combining it with a gauge theory of gravity, the so-called Cartan formalism,
we develop the Keldysh$+$Cartan formalism especially to calculate the HM and THC.
Although we can apply this formalism to disordered or interacting systems,
here we focus on the clean and non-interacting limit to reproduce the Wiedemann-Franz law.

This paper is organized as follows.
In Sect.~\ref{sec:grav}, we explain Luttinger's idea and revisit it from the gauge-theoretical viewpoint.
Based on this discussion, in Sect.~\ref{sec:gauge}, the HM and the thermal conductivity are defined as torsional responses, which is the first part of our results.
In Sects.~\ref{sec:keldysh} and \ref{sec:torsion}, we show the second part of our results,
i.e., the Keldysh$+$Cartan formalism and the first-order perturbation theory with respect to torsion.
We explicitly calculate the HM and the Kubo-formula contribution to the THC in Sects.~\ref{sec:hm} and \ref{sec:kappa}, respectively.
As a corollary, we discuss an effective action for the quantized THE in $(2 + 1)$-D TSCs in Sect.~\ref{sec:topo}.
Section~\ref{sec:summary} is devoted to the summary.

Let us summarize our notations.
We assign the Latin ($a, b, \ldots = {\hat 0}, {\hat 1}, \ldots, {\hat d}$) and Greek ($\mu, \nu, \ldots = 0, 1, \ldots, d$) alphabets to
locally flat coordinates and global coordinates, respectively.
We follow the Einstein convention, which implies summation over the spacetime dimension $D = d + 1$
when an index appears twice in a single term.
The Minkowski metric in a flat spacetime is taken as $\eta_{ab} = \diag(-1, +1, \ldots, +1)$.
The Planck constant and the electric charge are written as $\hbar$ and $q$, respectively.
We use $c = k_{\rm B} = 1$.
Upper or lower signs in equations correspond to bosons or fermions.

\section{Heat transport and gravity} \label{sec:grav}
First, we intuitively review Luttinger's idea that relates a gravitational potential to non-uniform temperature~\cite{PhysRev.135.A1505}.
We begin with an unperturbed system with a non-uniform chemical potential $\mu({\vec x})$ and temperature $T({\vec x}) = \beta^{-1}({\vec x})$.
The partition function in the local equilibrium is given by
\begin{equation}
  Z
  = \tr \exp \left[-\int {\rm d}^d x \beta({\vec x}) (H({\vec x}) - \mu({\vec x}) N({\vec x}))\right]. \label{eq:part1}
\end{equation}
The chemical potential and temperature are statistical forces and hence cannot be treated as perturbations.
Instead, we introduce mechanical forces $\phi({\vec x})$ and $\gamma({\vec x})$ by
\begin{subequations} \begin{align}
  \mu({\vec x})
  =& \gamma^{-1}({\vec x}) (\mu_0 - q \phi({\vec x})), \label{eq:phi} \\
  \beta({\vec x})
  = & \beta_0 \gamma({\vec x}), \label{eq:gamma} \\
  H^{\phi, \gamma}({\vec x})
  = & \gamma({\vec x}) H({\vec x}) + q \phi({\vec x}) N({\vec x}). \label{eq:ham}
\end{align} \label{eq:mech} \end{subequations}
Now the system is equivalent to the perturbed system in the uniform chemical potential $\mu_0$ and temperature $T_0 = \beta_0^{-1}$,
whose partition function is given by
\begin{equation}
  Z
  = \tr \exp \left[-\int {\rm d}^d x \beta_0 (H^{\phi, \gamma}({\vec x}) - \mu_0 N({\vec x}))\right]. \label{eq:part2}
\end{equation}
As a scalar potential $\phi({\vec x})$ is a mechanical force equivalent to the non-uniform chemical potential,
a gravitational potential $\gamma({\vec x})$, or $\phi_{\rm g}({\vec x}) = \gamma({\vec x}) - 1$, is equivalent to the non-uniform temperature.
However, it is still unclear why $\gamma({\vec x})$ is called a gravitational potential.

To answer this question, we revisit the fundamentals of heat transport, as well as charge transport, in terms of symmetries and gauge fields.
The Noether theorem tells us that a global continuous symmetry gives rise to a conservation law~\cite{9789812791719}.
The U($1$) gauge symmetry leads to the charge conservation law.
We can explicitly construct the charge current conserved when the on-shell condition is satisfied.
Then, according to the gauge principle, we require the local gauge symmetry.
Although a matter field alone does not possess it, the total system consisting of a matter field and a vector potential does.
This vector potential $A_i$,
which is introduced by replacing the partial derivative $\partial_{\mu}$ with the covariant one $\partial_{\mu} - i q A_{\mu}/\hbar$, is coupled to the charge current.

Such a discussion holds for heat transport.
The time translation symmetry gives rise to the energy conservation law.
Similarly, the space translation symmetry, which is usually absent in condensed-matter physics, leads to the momentum conservation law.
If we require the local spacetime translation symmetry, a corresponding gauge field is naturally introduced.
Since local spacetime translations are general coordinate transformations $x^{\mu} \to x^{\prime \mu}$,
the partial derivative transforms as a covariant vector, and the covariant derivative should be given by $h_a^{\phantom{a} \mu} \partial_{\mu}$.
This gauge field $h^a_{\phantom{a} \mu}$, called a vielbein, is coupled to the energy-momentum tensor.
In particular, $h^{\hat 0}_{\phantom{\hat 0} 0}$ is coupled to the Hamiltonian density
and hence is a gravitational potential that Luttinger introduced~\cite{PhysRev.135.A1505},
while $h^{\hat 0}_{\phantom{\hat 0} i}$ is coupled to the energy current defined by the product of the time derivative and the velocity.
Note that the other energy current defined by the product of the Hamiltonian density and the velocity
is obtained by imposing the on-shell condition but has no corresponding gauge field.

Now we can relate a theory of heat transport to gravity.
The local spacetime translation symmetry is required by the general covariance principle,
and a vielbein can be found in a gauge theory of gravity, i.e., the Cartan formalism~\cite{9789812791719}.
This formalism consists of two gauge fields;
a vielbein $h^a_{\phantom{a} \mu}$ and a spin connection $\omega^{ab}_{\phantom{ab} \mu} = -\omega^{ba}_{\phantom{ba} \mu}$.
The latter is associated with the local Lorentz symmetry and hence appears together with the generator of Lorentz transformations $S_{ab} = -S_{ba}$.
In these gauge fields, we have to replace the partial derivative $\partial_a$ with the covariant one,
\begin{equation}
  D_a
  \equiv h_a^{\phantom{a} \mu} (\partial_{\mu} - i q A_{\mu}/\hbar - i \omega^{ab}_{\phantom{ab} \mu} S_{ab}/2 \hbar). \label{eq:cov}
\end{equation}

More concretely, let us concentrate on a Dirac fermion.
The Dirac Lagrangian density in a curved spacetime is given by
\begin{equation}
  h L
  = h {\bar \psi} (\hbar \gamma^a D_a - m) \psi, \label{eq:dirac}
\end{equation}
in which $h = \det h^a_{\phantom{a} \mu}$ is the determinant of a vielbein,
${\bar \psi} \equiv i^{-1} \psi^{\dag} \gamma^{\hat 0}$ is the Dirac conjugate,
and the $\gamma$ matrices satisfy the Clifford algebra $\{\gamma^a, \gamma^b\} = 2 \eta^{ab}$.
First, we put $h^a_{\phantom{a} \mu} = \diag(\gamma, +1, \ldots, +1)$ to obtain
\begin{equation}
  h L
  = \psi^{\dag} [i \hbar \partial_0 - \gamma (-i \hbar {\vec \alpha} \cdot {\vec \nabla} + m \beta)] \psi. \label{eq:dirac1}
\end{equation}
Here ${\vec \alpha}$ and $\beta$ are the Dirac matrices,
and ${\vec \nabla}$ is not the covariant derivative in general relativity but the nabla in vector analysis.
Thus $h^{\hat 0}_{\phantom{\hat 0} 0} = \gamma$ is a gravitational potential coupled to the Hamiltonian density.
On the other hand, let us introduce an off-diagonal component $h^{\hat 0}_{\phantom{\hat 0} i} = A_{{\rm g} i}$.
Since the inverse has $h_{\hat \imath}^{\phantom{\hat \imath} 0} =  -A_{{\rm g} {\hat \imath}}$, Eq.~\eqref{eq:dirac} becomes
\begin{equation}
  h L
  = \psi^{\dag} [i \hbar \partial_0 - (-i \hbar {\vec \alpha} \cdot {\vec \nabla} + m \beta)
  - i \hbar {\vec \alpha} \cdot {\vec A}_{\rm g} \partial_0] \psi. \label{eq:dirac2}
\end{equation}
As expected, $h^{\hat 0}_{\phantom{\hat 0} i} = A_{{\rm g} i}$ is a ``vector potential''
coupled to the energy current defined by the product of the time derivative and the velocity ${\vec \alpha}$.

\section{Heat transport and torsion} \label{sec:gauge}
Next, we define the thermal conductivity and the HM based on the above discussion.
Since a vielbein and a spin connection are gauge fields, they induce field strengths called torsion and Riemann tensors:
\begin{subequations} \begin{align}
  T^a_{\phantom{a} \mu \nu}
  = & \partial_{\mu} h^a_{\phantom{a} \nu} + \omega^a_{\phantom{a} b \mu} h^b_{\phantom{b} \nu} - (\mu \leftrightarrow \nu), \label{eq:torsion2} \\
  R^a_{\phantom{a} b \mu \nu}
  = & \partial_{\mu} \omega^a_{\phantom{a} b \nu} + \omega^a_{\phantom{a} c \mu} \omega^c_{\phantom{c} b \nu}
  - (\mu \leftrightarrow \nu). \label{eq:riemann2}
\end{align} \label{eq:field2} \end{subequations}
If we choose $\omega^{ab}_{\phantom{ab} \mu} = 0$,
a torsional electric field $T^{\hat 0}_{\phantom{\hat 0} j0}$ is the mechanical force equivalent to a temperature gradient,
while a torsional magnetic field is the field strength coupled to the HM.
Thus the Kubo-formula contribution to the thermal conductivity and the HM are naturally defined by
\begin{subequations} \begin{align}
  T_0 {\tilde \kappa}^{{\hat \imath} {\hat \jmath}}
  \equiv & \frac{\partial J_{\rm Q}^{\hat \imath}}{\partial (-T^{\hat 0}_{\phantom{\hat 0} {\hat \jmath} {\hat 0}})}, \label{eq:defkappa} \\
  \beta_0 M_{{\rm Q} {\hat k}}
  \equiv & -\frac{1}{2} \epsilon_{{\hat \imath} {\hat \jmath} {\hat k}}
  \frac{\partial \Omega}{\partial (-\beta_0^{-1} T^{\hat 0}_{\phantom{\hat 0} {\hat \imath} {\hat \jmath}})}, \label{eq:defhm}
\end{align} \label{eq:def} \end{subequations}
where $J_{\rm Q}^{\hat \imath}$ is the heat current defined by the product of the time derivative and the velocity as discussed above,
and $\Omega \equiv E - T_0 S - \mu_0 N$ is the free energy.
This is the first part of our results and is justified below by explicitly deriving the Wiedemann-Franz law for the THC in the clean and non-interacting case.

Those in the fields of high-energy physics or cosmology may not be familiar with picking up a particular spacetime by hand.
In general relativity, we impose the torsion-free condition to determine the spin connection uniquely.
In this case, a temperature gradient is equivalent to the torsion-free spin connection $\omega_{{\hat \imath} {\hat 0} 0} = \partial_{\hat \imath} \gamma$.
However, as the electric conductivity is a current response to an electric field,
the thermal conductivity should be that to a field strength but not a gauge field.
Moreover, in condensed-matter physics where the Lorentz symmetry is usually absent, we cannot rely on a spin connection.
Therefore a torsional electric field is the best choice for a mechanical force equivalent to a temperature gradient.

Once a torsional electric field is fixed, it is not important to choose $h^{\hat 0}_{\phantom{\hat 0} 0}$ or $h^{\hat 0}_{\phantom{\hat 0} i}$.
Indeed, these two are connected by local time translations $x^{\prime 0} = x^0 + \xi^0(x)$.
Remember that a vielbein is a covariant vector
and transforms as $h^{\prime a}_{\phantom{\prime a} \mu} = h^a_{\phantom{a} \nu} \partial x^{\nu}/\partial x^{\prime \mu}$.
Even if we start from one spacetime with $h^{\hat0}_{\phantom{\hat 0} 0} = \gamma({\vec x})$, we can move to another spacetime with
\begin{subequations} \begin{align}
  h^{\prime {\hat 0}}_{\phantom{\prime {\hat 0}} 0}
  = (1 + \partial_0 \xi^0)^{-1} \gamma, \label{eq:timetr0} \\
  h^{\prime {\hat 0}}_{\phantom{\prime {\hat 0}} i}
  = -(1 + \partial_0 \xi^0)^{-1} \partial_i \xi^0 \gamma. \label{eq:timetr1}
\end{align} \label{eq:timetr} \end{subequations}
In particular, we can choose $h^{\prime {\hat 0}}_{\phantom{\prime {\hat 0}} 0} = 1$
and $h^{\prime {\hat 0}}_{\phantom{\prime {\hat 0}} i} = -x^0 \partial_i \gamma$.
Of course, both spacetimes give the same torsional electric field $T^{\hat 0}_{\phantom{\hat 0} j0} = \partial_j \gamma$.

On the other hand, we emphasize that a torsional magnetic field is essentially different from an angular velocity of rotation.
The former is induced by a vielbein $h^{\hat 0}_{\phantom{\hat 0} i} = A_{{\rm g} i}$, leading to the non-trivial metric,
\begin{equation}
  {\rm d} s^2
  = -({\rm d} t + {\vec A}_{\rm g} \cdot {\rm d} {\vec x})^2 + {\rm d} {\vec x}^2, \label{eq:line2}
\end{equation}
while the latter is represented by
\begin{equation}
  {\rm d} s^2
  = -{\rm d} t^2 + ({\rm d} {\vec x} + {\vec \phi}_{\rm r} {\rm d} t)^2, \label{eq:line3}
\end{equation}
with ${\vec \phi}_{\rm r} = {\vec \Omega} \times {\vec x}$.
Here ${\vec \Omega}$ is an angular velocity.
This metric is represented by a vielbein $h^{\hat \imath}_{\phantom{\hat \imath} 0} = \phi_{\rm r}^{\hat \imath}$.
These two vielbeins cannot be connected by local time translations but by local Lorentz transformations.
Nonetheless, these two metrics coincide in the gravito-electromagnetism, namely, when the second-order perturbations to the Minkowski metric are neglected.
It is a natural question whether these second-order perturbations are important even if linear responses are concerned.
From the gauge-theoretical viewpoint, a vielbein is primary, and a metric is secondary.
Since the latter is given by the square of the former, i.e., $g_{\mu \nu} = \eta_{ab} h^a_{\phantom{a} \mu} h^b_{\phantom{b} \nu}$,
we cannot drop these second-order perturbations in the metric formalism.
In other words, the gravito-electromagnetism is not even an approximation for either Eq.~\eqref{eq:line2} or Eq.~\eqref{eq:line3}.
This point was not correctly understood in the previous works~\cite{PhysRevB.85.045104,PhysRevLett.108.026802}.

Now an important problem is the experimental feasibility of a torsional magnetic field in condensed-matter physics.
In the context of high-energy physics and cosmology, this is closely related to the fundamental problem of `which action describes the correct gravity?'.
Remember that we use an electromagnet to generate a magnetic field in condensed-matter physics.
This is based on the Amp{\` e}re-Maxwell law, and, more fundamentally, on the Maxwell action in addition to the coupling between the charge current and a vector potential.
If there is the torsional analog of the Maxwell action, a torsional magnetic field is more or less feasible.
However, we cannot answer this problem because we only use torsion as external fields.

\section{Keldysh formalism in a curved spacetime} \label{sec:keldysh}
In this section, we present a general framework for calculating gravitational responses based on the Keldysh formalism.
We begin with the Keldysh formalism in a flat spacetime~\cite{9780521874991,9780521760829} to construct that in a curved spacetime.
The Dyson equation in a flat spacetime is well known,
\begin{equation}
  {\cal L}(x_1) {\hat G}(x_1, x_2) - {\hat \Sigma} \ast {\hat G}(x_1, x_2)
  = {\hat G}(x_1, x_2) {\cal L}(x_2) - {\hat G} \ast {\hat \Sigma}(x_1, x_2)
  = \delta(x_1 - x_2), \label{eq:dysonf}
\end{equation}
where ${\cal L}$ is the Lagrangian density, and $\ast$ is convolution.
The Keldysh Green function ${\hat G}$ and the self-energy ${\hat \Sigma}$ contain three independent real-time Green functions and self-energies, respectively.
We write them in the matrix form,
\begin{align}
  {\hat G}
  = & \begin{bmatrix}
        G^{\rm R} & 2 G^< \\
        0 & G^{\rm A}
      \end{bmatrix}, &
  {\hat \Sigma}
  = & \begin{bmatrix}
        \Sigma^{\rm R} & 2 \Sigma^< \\
        0 & \Sigma^{\rm A}
      \end{bmatrix}, \label{eq:matrix}
\end{align}
where ${\rm R}$, ${\rm A}$, and $<$ indicate the retarded, advanced, and lesser components defined by
\begin{subequations} \begin{align}
  G^{\rm R}(x_1, x_2)
  \equiv & -i \theta(t_1 - t_2) \left\langle [\psi(x_1), \psi^{\dag}(x_2)]_{\mp} \right\rangle, \label{eq:gr} \\
  G^{\rm A}(x_1, x_2)
  \equiv & +i \theta(t_2 - t_1) \left\langle [\psi(x_1), \psi^{\dag}(x_2)]_{\mp} \right\rangle, \label{eq:ga} \\
  G^<(x_1, x_2)
  \equiv & \mp i \left\langle \psi^{\dag}(x_2) \psi(x_1) \right\rangle. \label{eq:g<}
\end{align} \label{eq:green1} \end{subequations}
The lesser Green function acts as the density matrix and is useful for calculating thermal expectation values.

There are two effects of gravity.
One is to replace the volume element ${\rm d}^D x$ with the covariant one ${\rm d}^D x h(x)$.
Therefore the Dyson equation~\eqref{eq:dysonf} is modified by
\begin{equation}
  {\cal L}(x_1) {\hat G}(x_1, x_2) - {\hat \Sigma} \ast^h {\hat G}(x_1, x_2)
  = {\hat G}(x_1, x_2) {\cal L}(x_2) - {\hat G} \ast^h {\hat \Sigma}(x_1, x_2)
  = \delta^h(x_1, x_2), \label{eq:dysond1}
\end{equation}
where convolution and the $\delta$-function in a curved spacetime are defined by~\cite{9780521278584,9780521877879}
\begin{subequations} \begin{align}
  {\hat A} \ast^h {\hat B}(x_1, x_2)
  \equiv & \int {\rm d}^D x_3 h(x_3) {\hat A}(x_1, x_3) {\hat B}(x_3, x_2), \label{eq:convd} \\
  \delta^h(x_1, x_2)
  \equiv & h^{-1/2}(x_1) \delta(x_1 - x_2) h^{-1/2}(x_2). \label{eq:deltad}
\end{align} \label{eq:det} \end{subequations}
However, by introducing a tensor density,
\begin{equation}
  {\hb A}(x_1, x_2)
  \equiv h^{1/2}(x_1) {\hat A}(x_1, x_2) h^{1/2}(x_2), \label{eq:tensor}
\end{equation}
the Dyson equation can be written in the same form as that in a flat spacetime symbolically,
\begin{equation}
  {\cal L}(x_1) {\hb G}(x_1, x_2) - {\hb \Sigma} \ast {\hb G}(x_1, x_2)
  = {\hb G}(x_1, x_2) {\cal L}(x_2) - {\hb G} \ast {\hb \Sigma}(x_1, x_2)
  = \delta(x_1 - x_2). \label{eq:dysond2}
\end{equation}

In a flat spacetime, just symbolically, we can use the Wigner representation~\cite{9780521874991,9780521760829}.
This is a kind of Fourier transformation and makes it easy to deal with convolution.
We introduce the center-of-mass coordinate $X \equiv (x_1 + x_2)/2$ and the relative coordinate $x \equiv x_1 - x_2$
and then perform the Fourier transformation on the latter:
\begin{equation}
  {\hat A}(X, p)
  \equiv \int {\rm d}^D x e^{-i p_a x^a/\hbar} {\hat A}(X + x/2, X - x/2). \label{eq:wigner}
\end{equation}
The Wigner representation of convolution is given by the non-commutative Moyal product,
\begin{align}
  {\hat A} \ast {\hat B}(X, p)
  = & \int {\rm d}^D x \int {\rm d}^D y e^{-i p_a x^a/\hbar}  {\hat A}(X + x/2, y) {\hat B}(y, X - x/2) \notag \\
  = & \int {\rm d}^D x \int {\rm d}^D y \int \frac{{\rm d}^D q}{(2 \pi \hbar)^D} \int \frac{{\rm d}^D r}{(2 \pi \hbar)^D}
  e^{-i p_a x^a/\hbar} e^{i q_a (X - y + x/2)^a/\hbar} e^{i r_a (-X + y + x/2)^a/\hbar} \notag \\
  & \times {\hat A}(X/2 + y/2  +x/4, q) {\hat B}(X/2 + y/2 - x/4, r) \notag \\
  = & \int {\rm d}^D x_1 \int {\rm d}^D x_2 \int \frac{{\rm d}^D p_1}{(2 \pi \hbar)^D} \int \frac{{\rm d}^D p_2}{(2 \pi \hbar)^D} \notag \\
  & \times  e^{-i (p_{1 a}  x_2^a - p_{2 a} x_1^a)/\hbar} {\hat A}(X + x_1/2, p + p_1) {\hat B}(X + x_2/2, p + p_2) \notag \\
  = & \int {\rm d}^D x_1 \int {\rm d}^D x_2 \int \frac{{\rm d}^D p_1}{(2 \pi \hbar)^D} \int \frac{{\rm d}^D p_2}{(2 \pi \hbar)^D} \notag \\
  & \times [e^{-i p_{1 a} (x_2^a + i \hbar \partial_{p_a})/\hbar} e^{x_1^a \partial_{X^a}/2} {\hat A}(X, p)]
  [e^{i p_{2 a} (x_1^a - i \hbar \partial_{p_a})/\hbar} e^{x_2^a \partial_{X^a}/2} {\hat B}(X, p)] \notag \\
  = & {\hat A}(X, p) e^{i \hbar {\cal F}_0/2} {\hat B}(X, p). \label{eq:moyal}
\end{align}
We employ the inverse Fourier transformation in the second line
and change variables $x_1 = -X + y + x/2$, $x_2 = -X + y - x/2$, $p_1 = q - p$, and $p_2 = r - p$ in the third line.
The Poisson bracket in this phase space is defined by
\begin{equation}
  {\cal F}_0
  = \partial_{X^a} \otimes \partial_{p_a} - \partial_{p_a} \otimes \partial_{X^a}, \label{eq:poisson1}
\end{equation}
where a partial derivative on the left of $\otimes$ acts on the Wigner representation on the left-hand side, and vice versa.
We can expand the Moyal product with respect to $\hbar$ as $\ast = e^{i \hbar {\cal F}_0/2} = 1 + i \hbar {\cal F}_0/2 + \cdots$,
while we can construct the Moyal product from the Poisson bracket by using the deformation quantization~\cite{kontsevich2003}.
The commutation relation, which is assumed in the canonical quantization, can be derived by acting the Moyal product on $X^a$ and $p_b$.
In the Wigner representation, the Dyson equation~\eqref{eq:dysond2} becomes
\begin{equation}
  ({\cal L} - {\hb \Sigma}) \ast {\hb G}(X, p)
  = {\hb G} \ast ({\cal L} - {\hb \Sigma})(X, p)
  = 1. \label{eq:dysond3}
\end{equation}

Here we take into account the generator of Lorentz transformation $S_{ab}$,
which satisfies the Poincar{\' e} algebra, i.e., the commutation relations regarding $p_a$ and $S_{ab}$.
For those with a condensed-matter background, it might be better to call $S_{ab}$ spin.
To do that, we assume the extended Poisson bracket:
\begin{equation}
  {\cal P}_0
  \equiv (\partial_{X^c} \otimes \partial_{p_c} - \partial_{p_c} \otimes \partial_{X^c})
  + p_a \eta_{be} (\partial_{p_e} \otimes \partial_{S_{ab}} - \partial_{S_{ab}} \otimes \partial_{p_e})
  + \eta_{ac} S_{bd} \partial_{S_{ab}} \otimes \partial_{S_{cd}}. \label{eq:poisson2}
\end{equation}
The second and third terms describe the Poincar{\' e} algebra.
Such an extension was already done in the context of the twisted spin~\cite{Sugimoto15062012}.
Although the corresponding Moyal product $\ast$ may not be represented by such a simple form as $e^{i \hbar {\cal P}_0/2}$,
it is expanded with respect to $\hbar$ as $\ast = 1 + i \hbar {\cal P}_0/2 + \cdots$.
By using the Moyal product, if obtained, the Dyson equation~\eqref{eq:dysond3} is also extended to
\begin{equation}
  ({\cal L} - {\hb \Sigma}) \ast {\hb G}(X, p, S)
  = {\hb G} \ast ({\cal L} - {\hb \Sigma})(X, p, S)
  = 1. \label{eq:dysond4}
\end{equation}

The other effect of gravity is to replace the partial derivative $\partial_a$ with the covariant one in Eq.~\eqref{eq:cov}~\cite{9789812791719}.
This is realized by replacing the momentum $p_a$ with the mechanical momentum corresponding to the covariant derivative:
\begin{equation}
  \pi_a(X, p, S)
  = h_a^{\phantom{a} \mu}(X) (p_{\mu} - q A_{\mu}(X) - \omega^{ab}_{\phantom{ab} \mu}(X) S_{ab}/2). \label{eq:pi}
\end{equation}
Therefore the Dyson equation in gauge fields is represented by
\begin{equation}
  ({\cal L} - {\hb \Sigma}) \ast {\hb G}(X, \pi(X, p, S), S)
  = {\hb G} \ast ({\cal L} - {\hb \Sigma})(X, \pi(X, p, S), S)
  = 1. \label{eq:dysonc1}
\end{equation}
However, since $\pi_a(X, p, S)$ is a complicated function of $(X, p, S)$, it is convenient to change variables from $(X, p, S)$ to $(X, \pi, S)$.
This is a natural extension of the so-called Peierls substitution in the presence of a vector potential~\cite{PTP.116.61,PTP.117.415,Sugimoto15062012}.
Correspondingly, owing to the chain rule for partial derivatives, we obtain
\begin{subequations} \begin{align}
  \partial_{X^c}
  \to & h_c^{\phantom{c} \mu} \partial_{X^{\mu}} - h_c^{\phantom{c} \mu} h_d^{\phantom{d} \nu}
  (q \partial_{X^{\mu}} A_{\nu} + \partial_{X^{\mu}} h^a_{\phantom{a} \nu} \pi_a
  + \partial_{X^{\mu}} \omega^{ab}_{\phantom{ab} \nu} S_{ab}/2) \partial_{\pi_d}, \label{eq:chain1x} \\
  \partial_{\pi_c}
  \to & \partial_{\pi_c}, \label{eq:chain1p} \\
  \partial_{S_{ab}}
  \to & \partial_{S_{ab}} - \omega^{ab}_{\phantom{ab} \mu} h_c^{\phantom{c} \mu} \partial_{\pi_c}, \label{eq:chain1s}
\end{align} \label{eq:chain1} \end{subequations}
and then
\begin{subequations} \begin{align}
  \partial_{X^c} \otimes \partial_{p_c} -  \partial_{p_c} \otimes \partial_{X^c}
  \to & h_a^{\phantom{a} \mu} (\partial_{X^{\mu}} \otimes \partial_{\pi_a} - \partial_{\pi_a} \otimes \partial_{X^{\mu}}) \notag \\
  & + h_c^{\phantom{c} \mu} h_d^{\phantom{d} \nu}
  (q F_{\mu \nu} + t^a_{\phantom{a} \mu \nu} \pi_a + r^{ab}_{\phantom{ab} \mu \nu} S_{ab}/2) \partial_{\pi_c} \otimes \partial_{\pi_d}, \label{eq:chain2a} \\
  \pi_a \eta_{be} (\partial_{\pi_e} \otimes \partial_{S_{ab}} - \partial_{S_{ab}} \otimes \partial_{\pi_e})
  \to & \pi_a \eta_{be} (\partial_{\pi_e} \otimes \partial_{S_{ab}} - \partial_{S_{ab}} \otimes \partial_{\pi_e}) \notag \\
  & + h_c^{\phantom{c} \mu} h_d^{\phantom{d} \nu} (\omega^{ab}_{\phantom{ab} \mu} h_{b \nu} - \omega^{ab}_{\phantom{ab} \nu} h_{b \mu}) \pi_a
  \partial_{\pi_c} \otimes \partial_{\pi_d}, \label{eq:chain2b} \\
  \eta_{ad} S_{bc} \partial_{S_{ab}} \otimes \partial_{S_{cd}}
  \to & \eta_{ad} S_{bc} \partial_{S_{ab}} \otimes \partial_{S_{cd}}
  + \eta_{ad} S_{bc} \omega^{cd}_{\phantom{cd} e} (\partial_{\pi_e} \otimes \partial_{S_{ab}} - \partial_{S_{ab}} \otimes \partial_{\pi_e}) \notag \\
  & + h_c^{\phantom{c} \mu} h_d^{\phantom{d} \nu}
  (\omega^a_{\phantom{a} e \mu} \omega^{eb}_{\phantom{eb} \nu} - \omega^a_{\phantom{a} e \nu} \omega^{eb}_{\phantom{eb} \mu}) S_{ab}
  \partial_{\pi_c} \otimes \partial_{\pi_d}/2. \label{eq:chain2c}
\end{align} \label{eq:chain2} \end{subequations}
As a result, the Poisson bracket, Eq.~\eqref{eq:poisson2}, is perturbed as
\begin{align}
  {\cal P}
  = & h_a^{\phantom{a} \mu} (\partial_{X^{\mu}} \otimes \partial_{\pi_a} - \partial_{\pi_a} \otimes \partial_{X^{\mu}})
  + (\pi_a \eta_{be} + \eta_{ad} S_{bc} \omega^{cd}_{\phantom{cd} e})
  (\partial_{\pi_e} \otimes \partial_{S_{ab}} - \partial_{S_{ab}} \otimes \partial_{\pi_e}) \notag \\
  & + \eta_{ad} S_{bc} \partial_{S_{ab}} \otimes \partial_{S_{cd}}
  + (q T^a_{\phantom{a} cd} + T^a_{\phantom{a} cd} \pi_a + R^{ab}_{\phantom{ab} cd} S_{ab}/2) \partial_{\pi_c} \otimes \partial_{\pi_d}. \label{eq:poissonc}
\end{align}
The fourth term describes the emergent commutation relation between the mechanical momenta in the presence of gauge fields.
The corresponding Moyal product $\star$, which now we call the star product, is expanded with respect to $\hbar$ as $\star = 1 + i \hbar {\cal P}/2 + \cdots$.
It is an important future problem to construct this star product from the Poisson bracket by using the deformation quantization~\cite{kontsevich2003},
which helps us go beyond the first-order perturbation theory with respect to field strengths.
Note that the star product was already constructed in the presence of electromagnetic fields alone~\cite{PTP.117.415,Sugimoto15062012}.
By using the star product, the Dyson equation~\eqref{eq:dysonc1} attains the following simple form:
\begin{equation}
  ({\cal L}- {\hb \Sigma}) \star {\hb G}(X, \pi, S)
  = {\hb G} \star ({\cal L}- {\hb \Sigma})(X, \pi, S)
  = 1. \label{eq:dysonc2}
\end{equation}
The set of Eqs.~\eqref{eq:poissonc} and \eqref{eq:dysonc2} is the second part of our results.
It is a natural extension of a general framework for calculating electromagnetic responses previously established~\cite{PTP.116.61,PTP.117.415,Sugimoto15062012}
and enables us to calculate gravitational responses, as shown below.

\section{Perturbation theory with respect to torsion} \label{sec:torsion}
Let us derive the first-order perturbation theory with respect to the static and uniform torsion.
We drop the $X$-dependence in the Green function and the self-energy and impose the $S$-dependence on the band indexes.
We expand the star product, the Green function, and the self-energy as
\begin{subequations} \begin{align}
  \star
  = & 1 + i \hbar T^a_{\phantom{a} cd} \pi_a \partial_{\pi_c} \otimes \partial_{\pi_d}/2 + \cdots, \label{eq:startem} \\
  {\hb G}
  = & {\hat G}_0 + \hbar T^a_{\phantom{a} cd} {\hat G}_{T^a_{\phantom{a} cd}}/2 + \cdots, \label{eq:gtem1} \\
  {\hb \Sigma}
  = & {\hat \Sigma}_0 + \hbar T^a_{\phantom{a} cd} {\hat \Sigma}_{T^a_{\phantom{a} cd}}/2 + \cdots. \label{eq:stem}
\end{align} \label{eq:perturbtem} \end{subequations}
The Green function and the self-energy with the subscript $0$ indicate those in equilibrium,
and the Green function with the capital letter $G$ indicates that disorder and/or interactions are taken into account.
Note that Eq.~\eqref{eq:startem} is not semiclassical but exact and fully quantum-mechanical up to the first order with respect to the static and uniform torsion.
According to the deformation quantization, the exact form of the star product is defined by the infinite-order expansion with respect to $\hbar$
and may be represented by a couple of exponentials of the Poisson bracket~\cite{PTP.117.415,Sugimoto15062012}.
In general, even if we restrict ourselves to the first order with respect to torsion,
infinite terms containing the spacetime derivatives of torsion arise from a combination of the first and fourth terms in Eq.~\eqref{eq:poissonc}.
However, if torsion is static and uniform, we do not suffer from this infinite-order problem with respect to $\hbar$.

Below we calculate the first-order Green function with respect to torsion, i.e., ${\hat G}_{T^a_{\phantom{a} cd}}$.
This is totally in parallel with the previous calculation for that with respect to electromagnetic fields~\cite{PTP.116.61}
because the only difference is that torsion is coupled to $\pi_a$ while electromagnetic fields are coupled to $q$ as seen in Eq.~\eqref{eq:poissonc}.
By substituting Eq.~\eqref{eq:perturbtem} into the Dyson equation~\eqref{eq:dysonc2},
we obtain ${\hat G}_0 = ({\cal L} - {\hat \Sigma}_0)^{-1}$ and
\begin{equation}
  {\hat G}_{T^a_{\phantom{a} cd}}
  = {\hat G}_0 {\hat \Sigma}_{T^a_{\phantom{a} cd}} {\hat G}_0
  - \pi_a ({\hat G}_0 \partial_{\pi_c} {\hat G}_0^{-1} {\hat G}_0 \partial_{\pi_d} {\hat G}_0^{-1} {\hat G}_0 - (c \leftrightarrow d))/2 i. \label{eq:gtem2}
\end{equation}
In order to decompose this Keldysh Green function into three real-time Green functions, we use
\begin{subequations} \begin{align}
  ({\hat A}_1 \ldots {\hat A}_n)^{\rm R}
  = & A_1^{\rm R} \ldots A_n^{\rm R}, \label{eq:procr} \\
  ({\hat A}_1 \ldots {\hat A}_n)^{\rm A}
  = & A_1^{\rm A} \ldots A_n^{\rm A}, \label{eq:proca} \\
  ({\hat A}_1 \ldots {\hat A}_n)^<
  = & \sum_{i = 1}^n A_1^{\rm R} \ldots A_{i - 1}^{\rm R} A_i^<  A_{i + 1}^{\rm A} \ldots A_n^{\rm A}. \label{eq:proc<}
\end{align} \label{eq:proc} \end{subequations}
Furthermore, to calculate the lesser Green function,
we use the equilibrium condition $G_0^< = \pm (G_0^{\rm R} - G_0^{\rm A}) f(-\pi_{\hat 0})$ and introduce
\begin{subequations} \begin{align}
  G_{T^a_{\phantom{a} cd}}^<
  = & G_{T^a_{\phantom{a} cd}}^{< (0)} f(-\pi_{\hat 0}) + G_{T^a_{\phantom{a} cd}}^{< (1)} f^{\prime}(-\pi_{\hat 0}), \label{eq:gtem<1} \\
  \Sigma_{T^a_{\phantom{a} cd}}^<
  = & \Sigma_{T^a_{\phantom{a} cd}}^{< (0)} f(-\pi_{\hat 0}) + \Sigma_{T^a_{\phantom{a} cd}}^{< (1)} f^{\prime}(-\pi_{\hat 0}). \label{eq:stem<}
\end{align} \label{eq:tem<} \end{subequations}
Here $f(\epsilon) = (e^{\beta_0 \epsilon} \mp 1)^{-1}$ is the Bose or Fermi distribution function.
As a result, we obtain
\begin{subequations} \begin{align}
  G_{T^a_{\phantom{a} cd}}^{< (0)}
  = & \pm (G_{T^a_{\phantom{a} cd}}^{\rm R} - G_{T^a_{\phantom{a} cd}}^{\rm A})
  + G_0^{\rm R}
  [G_{T^a_{\phantom{a} cd}}^{< (0)} \mp (G_{T^a_{\phantom{a} cd}}^{\rm R} - G_{T^a_{\phantom{a} cd}}^{\rm A})] G_0^{\rm A}, \label{eq:gtem<2a} \\
  G_{T^a_{\phantom{a} cd}}^{< (1)}
  = & G_0^{\rm R} \Sigma_{T^a_{\phantom{a} cd}}^{< (1)} G_0^{\rm A} \notag \\
  & \mp \pi_a [(G_0^{\rm R} \partial_{\pi_c} G_0^{{\rm R} -1} (G_0^{\rm R} - G_0^{\rm A})
  - (G_0^{\rm R} - G_0^{\rm A}) \partial_{\pi_c} G_0^{{\rm A} -1} G_0^{\rm A}) \delta^d_{\phantom{d} {\hat 0}} - (c \leftrightarrow d)]/2 i, \label{eq:gtem<2b}
\end{align} \label{eq:gtem<2} \end{subequations}
and
\begin{subequations} \begin{align}
  G_{T^a_{\phantom{a} cd}}^{\rm R}
  = & G_0^{\rm R} \Sigma_{T^a_{\phantom{a} cd}}^{\rm R} G_0^{\rm R}
  - \pi_a (G_0^{\rm R} \partial_{\pi_c} G_0^{{\rm R} -1} G_0^{\rm R} \partial_{\pi_d} G_0^{{\rm R} -1} G_0^{\rm R} - (c \leftrightarrow d))/2 i, \label{eq:gtemr} \\
  G_{T^a_{\phantom{a} cd}}^{< (0)}
  = & \pm (G_{T^a_{\phantom{a} cd}}^{\rm R} - G_{T^a_{\phantom{a} cd}}^{\rm A}), \label{eq:gtem<0} \\
  \Sigma_{T^a_{\phantom{a} cd}}^{< (0)}
  = & \pm (\Sigma_{T^a_{\phantom{a} cd}}^{\rm R} - \Sigma_{T^a_{\phantom{a} cd}}^{\rm A}), \label{eq:stem<0} \\
  G_{T^a_{\phantom{a} {\hat \jmath} {\hat 0}}}^{< (1)}
  = & G_0^{\rm R} \Sigma_{T^a_{\phantom{a} {\hat \jmath} {\hat 0}}}^{< (1)} G_0^{\rm A}
  \mp \pi_a [G_0^{\rm R} \partial_{\pi_{\hat \jmath}} G_0^{{\rm R} -1} (G_0^{\rm R} - G_0^{\rm A})
  - (G_0^{\rm R} - G_0^{\rm A}) \partial_{\pi_{\hat \jmath}} G_0^{{\rm A} -1} G_0^{\rm A}]/2 i. \label{eq:gte<1}
\end{align} \label{eq:tem} \end{subequations}
The self-energies $\Sigma_{T^a_{\phantom{a} cd}}^{\rm R}$, $\Sigma_{T^a_{\phantom{a} cd}}^{\rm A}$,
and $\Sigma_{T^a_{\phantom{a} {\hat \jmath} {\hat 0}}}^{< (1)}$ are determined self-consistently, respectively,
and $G_{T^a_{\phantom{a} {\hat \imath} {\hat \jmath}}}^{< (1)} = \Sigma_{T^a_{\phantom{a} {\hat \imath} {\hat \jmath}}}^{< (1)} = 0$.

Before calculating the HM and the Kubo-formula contribution to the thermal conductivity,
let us comment on the relevance of this formalism compared to the previous theories on heat transport.
Reference~\cite{0022-3719-10-12-021} first revealed the necessity of the magnetization corrections
but involved the position operator, which is ill defined in periodic systems.
Reference~\cite{PhysRevLett.107.236601} overcame this problem and obtained the THC correctly,
but the scaling relations on the charge and heat currents were assumed without any microscopic explanations.
As a result, the non-trivial current corrections had to be calculated for each model.
On the other hand, our formalism can be applied to disordered or interacting systems without any assumptions or complicated calculations.
The former advantage was already shown
in the context of the anomalous Hall effect~\cite{PhysRevLett.97.126602,PhysRevB.77.165103,PhysRevB.78.041305,PhysRevB.79.195129}
and the orbital magnetization~\cite{PhysRevB.86.214415},
although we focus on the THE in the clean and non-interacting case below.
The Keldysh formalism in the presence of a gravitational potential alone is quite intriguing~\cite{PhysRevB.80.115111,PhysRevB.80.214516}
but is not sufficient for calculating the HM because a torsional magnetic field is not taken into account.

\section{Heat magnetization} \label{sec:hm}
In this section, we explicitly calculate the HM.
This is in parallel with the previous calculation for the orbital magnetization~\cite{PhysRevB.86.214415}.
Since it is difficult to calculate the proper HM defined by the free energy, let us calculate the auxiliary HM defined by the total energy.
In the Wigner representation, the total energy $K \equiv E - \mu_0 N$ is represented by
\begin{equation}
  K
  = \pm \frac{i \hbar}{2} \int \frac{{\rm d}^D \pi}{(2 \pi \hbar)^D} \tr [{\hb G} \star (-\pi_{\hat 0})]^< + \cc \label{eq:endens}
\end{equation}
Owing to the symmetrization, the star product is reduced to the ordinary product, and the auxiliary HM is given by
\begin{equation}
  {\tilde M}_{{\rm Q} {\hat k}}
  \equiv -\frac{1}{2} \epsilon_{{\hat \imath} {\hat \jmath} {\hat k}}
  \frac{\partial K}{\partial (-T^{\hat 0}_{\phantom{\hat 0} {\hat \imath} {\hat \jmath}})}
  = \pm \frac{i \hbar^2}{2} \epsilon_{{\hat \imath} {\hat \jmath} {\hat k}} \int \frac{{\rm d}^D \pi}{(2 \pi \hbar)^D} f(-\pi_{\hat 0}) (-\pi_{\hat 0})
  \tr G_{T^{\hat 0}_{\phantom{\hat 0} {\hat \imath} {\hat \jmath}}}^{< (0)}. \label{eq:auxhm1}
\end{equation}
Generally, the vertex corrections should be taken into account.
To translate this auxiliary HM to the proper HM, Eq.~\eqref{eq:defhm}, we have to solve the differential equation~\cite{PhysRevLett.107.236601}:
\begin{equation}
  \frac{\partial (\beta_0^2 M_{{\rm Q} {\hat k}})}{\partial \beta_0}
  = \beta_0 {\tilde M}_{{\rm Q} {\hat k}}. \label{eq:transhm}
\end{equation}
In practice, we can calculate the HM by using the set of Eqs.~\eqref{eq:tem}, \eqref{eq:auxhm1}, and \eqref{eq:transhm}.

Below, we restrict ourselves to the clean and non-interacting limit, i.e., ${\hb \Sigma} = 0$.
In this case, Eq.~\eqref{eq:auxhm1} is rewritten as
\begin{equation}
  {\tilde M}_{{\rm Q} {\hat k}}
  = \frac{\hbar^2}{2} \epsilon_{{\hat \imath} {\hat \jmath} {\hat k}} \int \frac{{\rm d}^D \pi}{(2 \pi \hbar)^D} f(-\pi_{\hat 0}) (-\pi_{\hat 0})^2
  \tr [g_0^{\rm R} v^{\hat \imath} g_0^{\rm R} v^{\hat \jmath} g_0^{\rm R} - ({\rm R} \to {\rm A})], \label{eq:auxhm2}
\end{equation}
where $g_0^{\rm R} = (-\pi_{\hat 0} - {\cal H} + \mu_0 + i \eta)^{-1}$ is the retarded Green function,
and $v^{\hat \imath} = -\partial_{\pi_{\hat \imath}} g_0^{{\rm R} -1}$ is the velocity.
By expanding the trace with respect to the Bloch basis that satisfies ${\cal H} | u_{n {\vec \pi}} \rangle = \epsilon_{n {\vec \pi}} | u_{n {\vec \pi}} \rangle$,
we obtain
\begin{align}
  {\tilde M}_{{\rm Q} {\hat k}}
  = & \frac{\hbar}{2} \epsilon_{{\hat \imath} {\hat \jmath} {\hat k}} \sum_{nm {\vec \pi}}
  \langle u_{n {\vec \pi}} | v^{\hat \imath} | u_{m {\vec \pi}} \rangle \langle u_{m {\vec \pi}} | v^{\hat \jmath} | u_{n {\vec \pi}} \rangle 
  \int \frac{{\rm d} \epsilon}{2 \pi} f(\epsilon) \epsilon^2 \notag \\
  & \times [(\epsilon - \epsilon_{n {\vec \pi}} + \mu_0 + i \eta)^{-2} (\epsilon - \epsilon_{m {\vec \pi}} + \mu_0 + i \eta)^{-1} - \cc] \notag \\
  = & \frac{i \hbar}{2} \epsilon_{{\hat \imath} {\hat \jmath} {\hat k}} \sum_{nm  {\vec \pi}}
  \frac{\langle u_{n {\vec \pi}} | v^{\hat \imath} | u_{m {\vec \pi}} \rangle \langle u_{m {\vec \pi}} | v^{\hat \jmath} | u_{n {\vec \pi}} \rangle}
  {(\epsilon_{n {\vec \pi}} - \epsilon_{m {\vec \pi}})^2} \notag \\
  & \times [2 f_{n {\vec \pi}} (\epsilon_{n {\vec \pi}} - \mu_0)^2
  - (2 f_{n {\vec \pi}} (\epsilon_{n {\vec \pi}} - \mu_0) + f^{\prime}_{n {\vec \pi}} (\epsilon_{n {\vec \pi}} - \mu_0)^2)
  (\epsilon_{n {\vec \pi}} - \epsilon_{m {\vec \pi}})] \notag \\
  = & \frac{1}{\hbar} \sum_{n {\vec \pi}}
  [\Omega_{n {\vec \pi} {\hat k}} f_{n {\vec \pi}} (\epsilon_{n {\vec \pi}} - \mu_0)^2
  - m_{n {\vec \pi} {\hat k}}
  (2 f_{n {\vec \pi}} (\epsilon_{n {\vec \pi}} - \mu_0) + f^{\prime}_{n {\vec \pi}} (\epsilon_{n {\vec \pi}} - \mu_0)^2)/2], \label{eq:auxhm3}
\end{align}
with $f_{n {\vec \pi}} \equiv f(\epsilon_{n {\vec \pi}} - \mu_0)$.
The Berry curvature $\Omega_{n {\vec \pi} {\hat k}}$ and the magnetic moment $m_{n {\vec \pi} {\hat k}}$ are defined by
\begin{subequations} \begin{align}
  \Omega_{n {\vec \pi} {\hat k}}
  \equiv & i \hbar^2 \epsilon_{{\hat \imath} {\hat \jmath} {\hat k}} \sum_m
 \frac{\langle u_{n {\vec \pi}} | v^{\hat \imath} | u_{m {\vec \pi}} \rangle \langle u_{m {\vec \pi}} | v^{\hat \jmath} | u_{n {\vec \pi}} \rangle}
  {(\epsilon_{n {\vec \pi}} - \epsilon_{m {\vec \pi}})^2}
  = i \hbar^2 \epsilon_{{\hat \imath} {\hat \jmath} {\hat k}} \langle \partial_{\pi_{\hat \imath}} u_{n {\vec \pi}} |
  \partial_{\pi_{\hat \jmath}} u_{n {\vec \pi}} \rangle, \label{eq:berrycurv} \\
  m_{n {\vec \pi} {\hat k}}
  \equiv & i \hbar^2 \epsilon_{{\hat \imath} {\hat \jmath} {\hat k}} \sum_m
 \frac{\langle u_{n {\vec \pi}} | v^{\hat \imath} | u_{m {\vec \pi}} \rangle \langle u_{m {\vec \pi}} | v^{\hat \jmath} | u_{n {\vec \pi}} \rangle}
  {\epsilon_{n {\vec \pi}} - \epsilon_{m {\vec \pi}}}
  = i \hbar^2 \epsilon_{{\hat \imath} {\hat \jmath} {\hat k}} \langle \partial_{\pi_{\hat \imath}} u_{n {\vec \pi}} |
  (\epsilon_{n {\vec \pi}} - {\cal H}) | \partial_{\pi_{\hat \jmath}} u_{n {\vec \pi}} \rangle. \label{eq:magmom}
\end{align} \label{eq:berry} \end{subequations}
We solve Eq.~\eqref{eq:transhm} to obtain the proper HM:
\begin{equation}
  M_{{\rm Q} {\hat k}}
  = -\frac{1}{\hbar} \sum_{n {\vec \pi}} 
  \left[\Omega_{n {\vec \pi} {\hat k}} \int_{\epsilon_{n {\vec \pi}} - \mu_0}^{\infty} {\rm d} z f(z) z
  + m_{n {\vec \pi} {\hat k}} f_{n {\vec \pi}} (\epsilon_{n {\vec \pi}} - \mu_0)/2\right]. \label{eq:hm}
\end{equation}

\section{Thermal conductivity} \label{sec:kappa}
Next we calculate the Kubo-formula contribution to the thermal conductivity.
In the Wigner representation, the thermal expectation value of the heat current is represented by
\begin{equation}
  J_{\rm Q}^{\hat \imath}
  = \pm \frac{i \hbar}{2} \int \frac{{\rm d}^D \pi}{(2 \pi \hbar)^D} \tr [v^{\hat \imath} \star {\hb G} \star (-\pi_{\hat 0})]^< + \cc, \label{eq:encurr}
\end{equation}
where $(-\pi_{\hat 0})$ is the Wigner representation of the covariant time derivative, and $v^{\hat \imath}$ is the renormalized velocity in general.
The Kubo-formula contribution to the thermal conductivity is defined in Eq.~\eqref{eq:defkappa} and calculated by
\begin{equation}
  T_0 {\tilde \kappa}^{{\hat \imath} {\hat \jmath}}
  = \mp i \hbar^2 \int \frac{{\rm d}^D \pi}{(2 \pi \hbar)^D} (-\pi_{\hat 0})
  \tr v^{\hat \imath} [G_{T^{\hat 0}_{\phantom{\hat 0} {\hat \jmath} {\hat 0}}}^{< (0)} f(-\pi_{\hat 0})
  + G_{T^{\hat 0}_{\phantom{\hat 0} {\hat \jmath} {\hat 0}}}^{< (1)} f^{\prime}(-\pi_{\hat 0})]. \label{eq:kappa1}
\end{equation}

We focus on the THC in the clean and non-interacting case, ${\hb \Sigma} = 0$.
In this case, Eq.~\eqref{eq:kappa1} is rewritten as
\begin{subequations} \begin{align}
  T_0 {\tilde \kappa}^{{\hat \imath} {\hat \jmath}}
  = & -\frac{\hbar^2}{2} \int \frac{{\rm d}^D \pi}{(2 \pi \hbar)^D} f(-\pi_{\hat 0}) (-\pi_{\hat 0})^2
  \tr [g_0^{\rm R} v^{\hat \imath} g_0^{\rm R} v^{\hat \jmath} g_0^{\rm R} - ({\rm R} \to {\rm A})] - (i \leftrightarrow j) \label{eq:kappa2a} \\
  & + \frac{\hbar^2}{2} \int \frac{{\rm d}^D \pi}{(2 \pi \hbar)^D} f^{\prime}(-\pi_{\hat 0}) (-\pi_{\hat 0})^2 \tr v^{\hat \imath}
  [g_0^{\rm R} v^{\hat \jmath} (g_0^{\rm R} - g_0^{\rm A}) - (g_0^{\rm R} - g_0^{\rm A}) v^{\hat \jmath} g_0^{\rm A}]. \label{eq:kappa2b}
\end{align} \label{eq:kappa2} \end{subequations}
The first term can be calculated as in Eq.~\eqref{eq:auxhm3}, and the second term is calculated as
\begin{align}
  \eqref{eq:kappa2b}
  = & \frac{\hbar}{2} \sum_{nm {\vec \pi}}
  \langle u_{n {\vec \pi}} | v^{\hat \imath} | u_{m {\vec \pi}} \rangle \langle u_{m {\vec \pi}} | v^{\hat \jmath} | u_{n {\vec \pi}}\rangle
  \int \frac{{\rm d} \epsilon}{2 \pi} f^{\prime}(\epsilon) \epsilon^2 \notag \\
  & \times [(\epsilon - \epsilon_{n {\vec \pi}} + \mu_0 + i \eta)^{-1} - (\epsilon - \epsilon_{n {\vec \pi}} + \mu_0 - i \eta)^{-1}]
  (\epsilon - \epsilon_{m {\vec \pi}} + \mu_0 + i \eta)^{-1} + \cc \notag \\
  = & -\frac{i \hbar}{2} \epsilon^{{\hat \imath} {\hat \jmath} {\hat k}} \sum_{nm {\vec \pi}}
  \frac{\langle u_{n {\vec \pi}} | v^{\hat \imath} | u_{m {\vec \pi}} \rangle \langle u_{m {\vec \pi}} | v^{\hat \jmath} | u_{n {\vec \pi}}\rangle - (i \leftrightarrow j)}
  {\epsilon_{n {\vec \pi}} - \epsilon_{m {\vec \pi}}} f^{\prime}_{n {\vec \pi}} (\epsilon_{n {\vec \pi}} - \mu_0)^2 \notag \\
  = & -\frac{1}{2 \hbar} \epsilon^{{\hat \imath} {\hat \jmath} {\hat k}} \sum_{n {\vec \pi}}
  m_{n {\vec \pi} {\hat k}} f^{\prime}_{n {\vec \pi}} (\epsilon_{n {\vec \pi}} - \mu_0)^2. \label{eq:kappa3}
\end{align}
In total, we obtain the Kubo-formula contribution to the THC:
\begin{align}
  T_0 {\tilde \kappa}^{{\hat \imath} {\hat \jmath}}
  = & -\frac{1}{\hbar} \epsilon^{{\hat \imath} {\hat \jmath} {\hat k}} \sum_{n {\vec \pi}}
  [\Omega_{n {\vec \pi} {\hat k}} f_{n {\vec \pi}} (\epsilon_{n {\vec \pi}} - \mu_0)^2
  - m_{n {\vec \pi} {\hat k}} (2 f_{n {\vec \pi}} (\epsilon_{n {\vec \pi}} - \mu_0) + f^{\prime}_{n {\vec \pi}} (\epsilon_{n {\vec \pi}} - \mu_0)^2)/2] \notag \\
  & - \frac{1}{2 \hbar} \epsilon^{{\hat \imath} {\hat \jmath} {\hat k}} \sum_{n {\vec \pi}}
  m_{n {\vec \pi} {\hat k}} f^{\prime}_{n {\vec \pi}} (\epsilon_{n {\vec \pi}} - \mu_0)^2 \notag \\
  = & -\frac{1}{\hbar} \epsilon^{{\hat \imath} {\hat \jmath} {\hat k}} \sum_{n {\vec \pi}}
  [\Omega_{n {\vec \pi} {\hat k}} (\epsilon_{n {\vec \pi}} - \mu_0)^2
  - m_{n {\vec \pi} {\hat k}} (\epsilon_{n {\vec \pi}} - \mu_0)] f_{n {\vec \pi}}. \label{eq:kappa4}
\end{align}
In the clean and non-interacting case, such terms involving $f^{\prime}_{n {\vec \pi}}$ are exactly canceled,
and the Kubo-formula contribution to the THC involves $f_{n {\vec \pi}}$ only.

It is known that the proper THC consists of the Kubo-formula contribution and the HM~\cite{0022-3719-10-12-021,PhysRevB.55.2344,PhysRevLett.107.236601}.
By combining Eqs.~\eqref{eq:kappa4} and \eqref{eq:hm}, we obtain the proper THC:
\begin{align}
  T_0 \kappa^{{\hat \imath} {\hat \jmath}}
  \equiv & T_0 {\tilde \kappa}^{{\hat \imath} {\hat \jmath}} + 2 \epsilon^{{\hat \imath} {\hat \jmath} {\hat k}} M_{{\rm Q} {\hat k}} \notag \\
  = & -\frac{1}{\hbar} \epsilon^{{\hat \imath} {\hat \jmath} {\hat k}} \sum_{n {\vec \pi}} \Omega_{n {\vec \pi} {\hat k}}
  \left[f_{n {\vec \pi}} (\epsilon_{n {\vec \pi}} - \mu_0)^2 + 2 \int_{\epsilon_{n {\vec \pi}} - \mu_0}^{\infty} {\rm d} z f(z) z\right]. \label{eq:kappa5}
\end{align}
Note that the Kubo-formula contribution, Eq.~\eqref{eq:kappa4}, and the HM, Eq.~\eqref{eq:hm}, are different from those in Ref.~\cite{PhysRevLett.107.236601},
but the proper THC, Eq.~\eqref{eq:kappa5}, is the same as that in Ref.~\cite{PhysRevLett.107.236601} and satisfies the Wiedemann-Franz law.
However, such a difference is not important because it is the proper THC only that can be measured in transport experiments.

The origins of the magnetization corrections can be understood as follows~\cite{PhysRevB.80.214516}.
In the Nernst effect, where the charge current flows perpendicular to a temperature gradient,
the orbital magnetization arises from a combination of the magnetization current ${\vec \nabla} \times {\vec M}$ and the $X$-dependence in the Green function.
On the other hand, in the Ettingshausen effect, where the heat current flows perpendicular to an electric field,
the magnetization energy ${\vec M} \cdot {\vec B}$ in the energy density translates into ${\vec E} \times {\vec M}$ in the energy current
through the Faraday law ${\vec \nabla} \times {\vec E} + {\dot B} = 0$.
In the THE, ${\vec M}$, ${\vec B}$, and ${\vec E}$ are replaced with the HM, a torsional magnetic field, and a torsional electric field, respectively.
Since the Faraday law originates from the Bianchi identity, it holds for torsional electromagnetic fields, too.
The coefficient $2$ in Eq.~\eqref{eq:kappa5} comes from these two mechanisms.

\section{Effective action for the quantized thermal Hall effect} \label{sec:topo}
As in the case of the Hall conductivity quantized in $(2 + 1)$-D time-reversal-broken topological insulators~\cite{RevModPhys.82.3045,RevModPhys.83.1057},
the THC is known to be quantized in $(2 + 1)$-D TSCs~\cite{PhysRevB.61.10267,PhysRevLett.108.026802,JPSJ.82.023602}.
On the basis of the gauge-theoretical viewpoint and the Widemann-Franz law, we discuss the effective action for this quantized THE.
First, let us reproduce the heat analog of the St{\v r}eda formula already obtained in Ref.~\cite{PhysRevLett.108.026802}.
From Eq.~\eqref{eq:hm}, we get
\begin{equation}
  T_0 \frac{\partial M_{{\rm Q} {\hat k}}}{\partial T_0}
  = -\frac{1}{\hbar} \sum_{n {\vec \pi}}
  \left[\Omega_{n {\vec \pi} {\hat k}} \int_{\epsilon_{n {\vec \pi}} - \mu_0}^{\infty} {\rm d} z (-f^{\prime}(z)) z^2
  - m_{n {\vec \pi} {\hat k}} f^{\prime}_{n {\vec \pi}} (\epsilon_{n {\vec \pi}} - \mu_0)^2/2\right], \label{eq:streda1}
\end{equation}
in which the first term is equal to Eq.~\eqref{eq:kappa5}, and the second term can be dropped in gapped systems at sufficiently low temperature.
As a result, the St{\v r}eda formula is given by~\cite{PhysRevLett.108.026802}
\begin{equation}
  \kappa^{{\hat \imath} {\hat \jmath}}
  = \epsilon^{{\hat \imath} {\hat \jmath} {\hat k}} \frac{\partial M_{{\rm Q} {\hat k}}}{\partial T_0}. \label{eq:streda2}
\end{equation}

In $(2 + 1)$-D gapped systems, the THC is quantized owing to the Wiedemann-Franz law~\cite{PhysRevB.61.10267,PhysRevLett.108.026802,JPSJ.82.023602},
\begin{equation}
  \kappa^{xy}
  = \frac{\pi^2 T_0}{3 q^2} \sigma^{xy}
  = -\frac{C \pi T_0}{6 \hbar}, \label{eq:qthe1}
\end{equation}
with $C$ being the first Chern number.
By using the St{\v r}eda formula, Eq.~\eqref{eq:streda2}, the temperature dependence of the HM at low temperature is given by
\begin{equation}
  M_{{\rm Q} z}
  = M_{{\rm Q} z}(T_0 = 0) - \frac{C \pi T_0^2}{12 \hbar}. \label{eq:qthe2}
\end{equation}
Apart from the ground-state value $M_{{\rm Q} z}(T_0 = 0)$, Eqs.~\eqref{eq:qthe1} and \eqref{eq:qthe2} are described by the effective action:
\begin{equation}
  S_{\rm eff}
  = -\frac{C \pi}{24 \hbar} \int {\rm d}^3 X
  \epsilon^{\mu \nu \lambda} (T h^{\hat 0}_{\phantom{\hat 0} \mu}) \partial_{X^{\nu}} (T h^{\hat 0}_{\phantom{\hat 0} \lambda}). \label{eq:tcs1}
\end{equation}
This is a corollary of the gauge-theoretical discussion above.
If the heat current is defined by the product of the Hamiltonian density and the velocity, which has no corresponding gauge field,
we can obtain the Wiedemann-Franz law and the St{\v r}eda formula~\cite{PhysRevLett.107.236601,PhysRevLett.108.026802}
but not this effective action.
In the case of TSCs, an extra factor $1/2$ should be multiplied owing to their Majorana nature~\cite{PhysRevLett.108.026802}.
Equation~\eqref{eq:tcs1} is the Lorentz-temporal part of the torsional Chern-Simons action,
\begin{equation}
  S_{\rm tCS}
  = \frac{\hbar}{4 \pi l^2} \int {\rm d}^3 X
  \eta_{ab} \epsilon^{\mu \nu \lambda} h^a_{\phantom{a} \mu}
  (\partial_{X^{\nu}} h^b_{\phantom{b} \lambda} + \omega^b_{\phantom{b} c \nu} h^c_{\phantom{c} \lambda}), \label{eq:tcs2}
\end{equation}
with $l^{-1} \propto T$.
Note that the Lorentz-spatial part describes the topological Hall viscosity but has different $l^{-1}$~\cite{PhysRevLett.107.075502,Hidaka01012013,PhysRevD.88.025040}.
The temporal and spatial parameters do not necessarily coincide because we do not assume the Lorentz symmetry.

\section{Summary} \label{sec:summary}
To summarize, we have revisited a theory of heat transport from the gauge-theoretical viewpoint of gravity
and defined the HM and the Kubo-formula contribution to the thermal conductivity as torsional responses.
In addition, we have developed the Keldysh$+$Cartan formalism to calculate these quantities without any unfounded assumptions or complicated calculations.
This is a natural extension of the gauge-covariant Keldysh formalism for calculating electromagnetic responses
and can be easily applied to disordered or interacting systems.
We have reproduced the THC satisfying the Wiedemann-Franz law in the clean and non-interacting case.
We have also discussed the effective action for the quantized THE in $(2 + 1)$-D TSCs.

{\it Note added.} After the first submission of this paper, we wrote a related paper on heat polarization~\cite{JPSJ.83.033708}.
Although polarization is coupled to a kind of electric field, it cannot be defined in periodic systems by the electric-field derivative of the free energy.
We employed the gradient expansion by using the first term in the Poisson bracket, Eq.~\eqref{eq:poissonc}, to define the heat polarization.
The $(3 + 1)$-D analog of Eq.~\eqref{eq:tcs1} was also discussed in the context of the effective action for the cross-correlation responses in $(3 + 1)$-D TSCs.

\section*{Acknowledgements} \label{sec:acknowledge}
We thank T. Kimura, N. Sugimoto, K. Shiozaki, and H. Sumiyoshi for fruitful discussions and S. Fujimoto for careful reading of this manuscript.
This work was supported by Grants-in-Aid for the Japan Society for the Promotion of Science Fellows No.~$24$-$600$.


\end{document}